\documentclass[12pt]{article}
\usepackage{epsfig}
\usepackage{rotate}
\begin{document}
\vspace*{3cm}
\begin{center}
{\Large\bf{THE POSSIBLE ROLE OF SALAM'S EFFECT IN FUNDAMENTAL INTERACTIONS
UNIFICATION}}\\[5mm]
{\large{R.G. Ragachurin}}\\[5mm]
{\small Institute of Nuclear Physics, Tashkent, Uzbekistan}
\end{center}

\section{SALAM'S RANGE}
\par The effect of the rest mass compensation (Salam's effect [1]) can
play an important role in the unification of the fundamental interactions
if one adresses to the field equation of the energy conservation. According
to [2], this equation is
\begin{equation}
\frac {\partial}{\partial t}\lbrace \int W dv + E_r \rbrace = - \oint \vec {f} d\vec {s},
\end{equation}
where $W$ is the density of the field energy, $E_r$ is the relativistic
kinetic energy of particles, $\vec f$ is Pointing's vector, $dv$
and $d\vec s$ are the elements of volume and surface respectively.
\par One of the important features of Salam's effect is connected with
the formation of a potential barrier around the particle whose rest
mass is spent on the barrier formation. This barrier is important
for the consideration of stationary solutions of equation (1), when
in the left hand side the time derivative from the sum should be zero.
This condition is satisfied if the surface integral in the right hand
side is equal to zero too that is possible in the case when ingoing
and outgoing energy fluxes are equal in the absolute quantity. The ingoing
flux is arising from an external field and outgoing - from Salam's
barrier. Due to their equality the stationary interaction of a particle
with an external field becomes possible. Thus, one of Salam's effect
features is connected with the maintenance of this stationary condition.
\par The influence of the effect can be performed in some range which
is determined by variations of the external field and the relativistic
kinetic energy $E_r$. The presence of the range borders is caused
by the limited rest mass of the particle. One of the borders arises
at zero influence of the external field (the particle infinitely far
from a field source) and minimum possible relativistic kinetic energy
$E_r$ (rest energy). There is no necessity for the barrier formation
in this case. The second border arises in the case when all rest mass
is spent on the maintenance of the stationary condition and further
increase of the external field and relativistic kinetic energy $E_r$
cannot results in the formation of the stationary states. The range
of the effect action is defined as Salam's range.
\vspace*{2cm}
\section{RELATIVITY AND GEOMETRIZATION PRINCIPLE}

\par The choice of the geometrization principle plays the important
role too. The choice is defined by the presence of the relativistic
kinetic energy $E_r$ in equation (1),
\begin{equation}
E_r = mv^2 [1 - (v/c)^2]^{-1/2},
\end{equation}
where $m$ is the rest mass of the particle, $v$ is its speed and $c$ is
the speed of light in vacuum.
\par The expansion of the relativistic kinetic energy $E_r$ in McLoren's
series on speed $v$ = 0 is
\begin{equation}
E_r = mc^2 + (1/2)mv^2 + (3/8)mv^4/c^2 + (5/16) mv^6/c^4 + ...
\end{equation}
\par There are only terms with even degrees of the ratio $(v^2/c^2)^n$ in
the equation (3). The absence of the odd terms is connected with the equality
of relevant derivatives in McLoren's series to zero.
\par The interesting feature of the equation (3) is connected with the
second term which defines the nonrelativistic kinetic energy $E_k$.
Higher degrees terms also have physical meaning of an energy but differ from
the nonrelativistic energy $E_k$. Their presence allows to use the
potential geometrization principle according to which changes of the field
potential energy should be equal to changes of the nonrelativistic
kinetic energy $E_k$. Then the sum
\begin{equation}
E_{r3} = (3/8)mv^4/c^2 + (5/16)mv^6/c^4 + ...
\end{equation}
has the important physical meaning. Our task is to show that the sum (4)
defines the energy of the gravitational interaction under special circumstances
to be specified.

\section{THE EFFECTIVE VOLUME OF SALAM'S BARRIER}

\par The role of Salam's barrier effective volume (which is defined
by the effective radius $r_{eff}$ in the spherical approximation)
is important at the consideration of the stationary solutions of the
equation (1).
\par Let us assume, that the interaction of the particle with the field
is executed inside Salam's range. Then the left hand side of the equation (1)
must be zero or
\begin{equation}
\int Wdv + E_r = Const.
\end{equation}
It is necessary to consider interaction conditions at the case of the
particle infinitely far from a source of the field, minimal possible
value of the relativistic kinetic energy $E_r$ and zero considered
volume. If the distance of the particle from the field source is determined
by the generalized coordinate $R$, the minimal value of the relativistic
kinetic energy $E_r$ - by $mc^2$ (the rest energy), then it follows
from the expression (5) that at the infinity $R_{\infty}$ and at
zero volume
$$
Const = mc^2
\nonumber\\
$$
and therefore
\begin{equation}
\int Wdv + E_r = mc^2.
\end{equation}
\par The equation (6) is defined as the stationary one which area
of the solutions is inside Salam's range. It is assumed that the variations
of the nonrelativistic kinetic energy $E_k$ are only connected with
the central interaction and do not depend on the integration volume.
\par It follows from the equation (6) that
\begin{equation}
E_r = mc^2
\end{equation}
at the coordinate $R_{\infty}$ and the zero volume. The last is the
first border of Salam's range.
\par Let the volume increases at the same value of the coordinate
$R_{\infty}$. The contribution to the space integral from the central
interaction remains is being equal to zero in this case. However,
this integral increases due to the non-central interaction. The area
of the field which is included in the space integral is defined as
the effective volume of Salam's barrier. The increase of the effective
volume (increase of $r_{eff}$) should be accompanied by the equivalent
decrease of the relativistic kinetic energy $E_r$ for the execution
of the equation (6). The limit is reached in the case when all rest
mass is transformed into the energy of the barrier. This case defines
the second border of the range.
\par The limiting states of the particle are represented by points
$a$ and $b$ in the figure 1. The energy stationary states are presented
there at various coordinates $R$ and $r_{eff}$. The longitudinal variations
are defined by $R$, and the lateral ones- by $r_{eff}$. The smooth curve
represents the field energy variations which are connected to the
potential energy $E_p$ with the negative sign. The field energy has
a positive value, and the potential energy $E_p$ has the negative
value in the connected systems. Therefore, the value $-E_p$ defines
the field energy variations corresponding to the variations of the
nonrelativistic kinetic energy $E_k$ in the connected systems. The
connected systems are considered below.
\par The limiting states a and b are defined by the conditions
$$
\left \{
\begin{array}{ l l }
E_r  =  mc^2 \ ,   \ \ \ \ \ \ \ \ \ \ \ & \nonumber\\
\int Wdv  =  0 \ ,\ \ \ \ \ \ \ \ \ \ \ \ \ \ \ \ \ \ \ & \nonumber \\
r_{eff} =    0 \ ;   \ \ \ \ \ \ \ \ \ \ (8a) &  \nonumber\\
\end{array}
\right.
$$
and
\begin{equation}
\left \{
\begin{array}{ l l }
E_r = 0 \ , \ \ \ \ \ \ \ \ \ \ \ & \nonumber \\
\int Wdv = mc^2 \ ,\ \ \ \ \ \ \ \ \ \ \ \ \ \ \ \ \ \ \ & \nonumber \\
r_{eff} = r_{effm} \ ,\ \ \ \ \ \ \ \ \ \ (8b) & \\
\end{array}
\right.
\end{equation}
where the value $r_{effm}$ corresponds to the transformation of all
rest mass into the barrier energy.
\par The borders of the range, similar to (8), exist at any value
of the coordinate $R$. Particular values $E_r$, $\int W dv$ and $r_{eff}$
vary in dependence on the sum of the series (3) but the physical meaning
of the borders is the same. The type (8a) borders define the maximal
mass for the concrete $R$, and the type (8b) borders - a state in
which this mass is completely transformed into the energy of the barrier.
\par The above mentioned discussion shows that the influence of Salam's
effect results in the additional coordinate $r_{eff}$ account. The
coordinate $r_{eff}$ is connected to the energy lateral variations as against
to the generalized coordinate $R$ which determines longitudinal ones.

\section{THE PHYSICAL MEANING OF THE POTENTIAL ENERGY}

\par The energies $\int W dv$ and $E_k$ grow with the decrease $R <
R_{\infty}$. The integrated field energy grows due to the density
$W$ increase. All terms of the series (3) with higher degrees should
grow too because of the nonrelativistic kinetic energy $E_k$ increase.
Therefore, the particle has to spend the whole rest mass for the increase
effects compensation because the series (3) and (4) are infinite ones.
As result, the exclusion of these series terms arises because of limiting
$m$. The sequence of the terms exception is essential during this
process.
\par The process of the exclusion is illustrated in the figure 1 by
any state corresponding to the coordinate $R_o$. The various segments
define the variations of the energy connecting to the different members
of the equation (6).
\par The segment $R_oa'$ is equal (relative to the value $R_{\infty}$)
to the variation of the field energy due to the central interaction.
It should be equal to the variation of the nonrelativistic kinetic
energy $E_k$ due to the geometrization principle action. This part
is connected with the contribution of the space integral into the
equation (6) and defines the scalar part of the interaction. This
part should be taken into account twice due to independent entry of
the space integral and the relativistic kinetic energy $E_r$ (due
to the second term of the expansion (3)). The repeated account is
represented by the segment $b'c$ which should be equal to the segment
$R_oa'$ on energy value but differs from it on the physical meaning.
The difference is connected with the fact that the quantity $R_oa'$
is defined by the variation of the longitudinal coordinate $R$, and
the quantity $b'c$ - by the variation of the lateral coordinate $r_{eff}$.
This part is connected with the vector interaction.
\par The particle should compensate the variations which are caused
by the terms of the infinite series (4) in addition to the compensation
of the energy variations which are connected to segments $R_oa'$ and
$b'c$. But, according to the equation (6), it only has an energy part
which is equal to the segment $a'b'$. The necessity of the exclusion
of the terms quantity arises there.
\par The part of the interaction connected with the sum of the series
(4) is defined as the tensor part of the interaction.
\par The division of the interaction on the scalar, vector and tensor
parts is connected with distinction in their functional role.
\par The scalar part is connected to the change of the field energy relative
to the coordinate $R_{\infty}$ and consequently defines that event
which have taken place during decrease of the coordinate $R$ before
appearance of the particle in the considered point of the space ($R_o$
is in the figure 1).
\par The vector part is connected to the lateral changes of the effective
radius $r_{eff}$. Analogously to the scalar part, it is connected
with the event which has happened before appearance in the considered
point. However, it limits Salam's range on the part of the smaller values
of the rest mass on the contrary to the scalar part.
\par The functional role of the tensor part is connected with the
cross changes of the effective radius $r_{eff}$. The meaning of its
allocation in the separate part is connected with its potential influence.
One can easily see that the following expression
\begin{equation}
mv^2 = mc^2 - E_{r3}.
\end{equation}
is true for any value of the coordinate $R$.
\par It follows from the expression (9) that the increase of the quantity
$mv^2$ occurs due to the decrease of the tensor part $E_{r3}$. Therefore,
the quantity $E_{pl}$ defined from the equality
\begin{equation}
E_{pl} = - E_{r3}
\end{equation}
is responsible in fact for the potential energy of the field. In any
space point $R_o$ this energy has the meaning of the energy stock
which will be transformed into the nonrelativistic kinetic energy
at the decrease $R<R_o$. Thus, the functional division of the interaction
on three parts in the lateral direction leads to the division of the
space in the longitudinal one.
\par The action of the potential energy $E_{pl}$ has an unexplicit
character. It leads to the compensation of all members of the series
(4). Therefore, the particle has only the scalar and vector explicit
components of the energy.
\par All terms of the energy $E_{r3}$ increase at the decrease $R<R_{\infty}$.
Therefore, the summary decrease of this energy [which is necessary
for the increase of the energy $mv^2$ in the expression (8)] can be
executed only due to the decrease (relative to $R_{\infty}$) of the
terms quantity in the series (4). The process of the decrease is connected
with the formal part of the series (3). This series is obtained as
a result of the expansion in McLoren's series which is connected to
the definition of the derivatives $f(0)$, $f'(o)$, $f''(0)$ and etc.
One can easily see that the double mechanism of the series terms exclusion
arises at the calculations of the derivatives $f^{2k}(0)$. It is connected
with the fact that the series (3) is power series in which the terms
are located on the growing degrees of $(v^2/c^2)^n$. When $k < n$,
the terms in $f^{2k}(0)$ are excluded on the part of the small degrees
(because all of them are equal to zero) due to this property. When
$k > n$, the terms are excluded on the part of the high degrees due
to the presence of the terms $\sim v^{2k}$. In result, there is only
one term (for which $k = n$) in $f^{2k}(0)$. The action of this double
mechanism results in the strict order of a position of the series
terms in the segment $a'b'$ - the first segment has to represent the
first term of the series (4), the second one - the second term and
so one. This feature is represented by the points 1 and 2 at the coordinate
$R_o$ in the figure 1. The segment $a'1$ represents the first term
of the series (4), $12$ - the second one and so on.
\par One can see that the whole segment $a'b'$ represents the discrete
sum of the segments with the strictly certain length for each term
and with the strict order of the position of all terms. This order
can not be changed without the special external influence on the closed
system (for example, to present a segment $a'1$ as the sum of the
several terms with higher degrees). The strict order of the position
leads to the integer number of the derivatives which enter into the
segment $a'b'$. The stationary equation (6) is satisfied only in this
case.
\par One can wait that the process of the decrease of the terms of
the series (4) is connected with the curvature of the space representation
of the Salam's barrier due to the strict order and integer number of
the derivatives. Then, the first stationary state is connected with
the exclusion of the first derivative (relative to $R_{\infty}$), the
second one - with the second derivative and so on.

\section{A CYCLE OF GAUGE TRANSFORMATIONS}

\par Some results presented here after can be obtained with the analysis
of the parallel displacement of the unit vector (first considered by
Weyl [3]). However, our results definitely show that the length of
the vector is meaning full. The correlation between pseudo-euclidean
and riemannian spaces (first mentioned in works [4-7]) has the important
meaning too. Besides, the model gives the clear physical understanding
of the phenomenon.
\par In the model the length of the vector is connected to the quantity of
the derivatives which are excluded from the series (4) at the transition
between two adjacent stationary states. The curvature characteristics
of Salam's barrier and the variation of the angular momentum do not
depend on the length of the vector. However, the functional completeness
of the stationary states is violated at such transition.
\par The model gauge transformations are based on the representation
of one cycle. This cycle is connected with the transition from scalar
to tensor spaces and back. The model gauge transformations are similar
to Weyl's ones.
\par The understanding of the physical meaning of the rest mass is
necessary for the elucidation of the role of the model gauge cycle.
One has to come back to consideration of the borders of Salam's range
at the coordinate $R_{\infty}$ (the section 3). This consideration
shows that the meaning of the field compressed in some curvilinear
volume can be attributed to the rest energy. The rest mass defines
the energy compression in the lateral direction at $R_{\infty}$. Such
object is retained in the stationary state by the equality of the
external field energy to internal one in its volume.
\par At the decrease $R<R_{\infty}$ the volumes of the Salam's barriers
are decreased. Therefore, the part of the internal energy become free
due to the volumes decrease. According to the previous section, this
part must be defined by the corresponding derivative of the series
(4). The energy of this part, which is defined by the lateral coordinate
$r_{eff}$, is transformed into the energy of variation of the longitudinal
coordinate $R$ (nonrelativistic kinetic energy $E_k$) at one gauge
cycle. Three various space areas (scalar, vector and tensor ones)
correlate at such cycle. The scalar area is represented by symbol
I, the vector - by symbol II and the tensor - by symbol III in the
figure 1.
\par The model cycle is defined by one elementary segment which is
excluded in the process of the decrease of the terms quantity of
the series (4). One elementary lateral segment is transformated in
a longitudinal one with help of such cycle. The scale of such transformation
is defined by the functional dependence $-E_p(R)$. The consequtive
recurrence of such cycles leads to the strict sequence of the longitudinal
elementary segments $\Delta R_i$ due to the strict sequence of the
lateral ones. In this sequence each segment in the scale $-E_p(R)$
reproduces the appropriate term of the series (4).
\par The considered mechanism has the important property - the curve
$-E_p$ plays a role of a string connecting all possible stationary
states inside Salam's range.
\par Some common conclusions can be received by the consideration
of one elementary cycle. Such cycle is represented by the dush line
$aR_oa'1d'$ in the figure 1. This line is formed as a result of three
transitions. The first transition occurs from the point $a$ which
is located on the curve $-E_p$ (the border between the scalar and
tensor areas) to a point $R_o$. The second transition occurs from
the point $R_o$ to the state $1$ and is connected with motion from
the scalar to the tensor area. At last, the third transition occurs
from the point $1$ to the point $d'$ before the crossing with the
curve $-E_p$. One can see that the first transition images on the
longitudinal direction in the scale $-E_p$ an interval $\Delta R$
corresponding to the second term of the series (3) - $mv^2/2$. The
second transition reproduces before the crossing with the curve $-E_p$
value $mv^2/2$, and after crossing up to the point $1$ - the value
of the first term of the series (4). At last, the third transition
images this term on the axis $R$ in the appropriate scale. One can
see that the consequtive reccurence of the gauge cycles results in
consequtive projection of all terms of the series (4).
\par All gauge cycles have three topological features which are important
for the further discussion: \\
1. All gauge cycles are formed with help of three transition connected
with three movements between scalar and tensor areas; \\
2. Considered values have the phase shift on an angle $\pi$ at the
tensor area (For example, the rest mass has a maximal value at the
bottom border; the spacing of the segments which define the value
of the nonrelativistic kinetic energy $E_k$ [relative to the segments
which define the corresponding variation of the coordinate $R$) has the
phase shift on the same angle relative to the scalar area etc.]. An
input from the scalar area to tensor one on the lateral part of the
cycle and then the output on the longitudinal part can be treated
as rotation of the system on an angle $2\pi$ due to this feature; \\
3. The point of the crossing with the curve $-E_p$ defines the begining
or end of an elementary segment on any transition. The states which
are located on the edges of the transition (for example, $R_o$ and
$1$ in the figure 1) can be considered as the states having the common
point on the curve $-E_p$ due to this property.

\section{THE ROLE OF THE MASSIVE PARTICLE UNLOCAL STRUCTURE IN
TWO-PARTICLES INTERACTION}

\par Three common properties which were considered in the previous
section are important in the consideration of two-particles interaction.
Such consideration requires explicit dependence of $W(R)$. According
to [2],
\begin{equation}
W = \frac{E^2 + H^2}{8\pi},
\end{equation}
where $E$ and $H$ are the intensity of electrical and magnetic fields,
respectively.
\par The magnetic field is assumed to be zero in our research, and
$E$ is defined by Coulomb interaction of two particles with electrical
charge $\pm e$ (hydrogen atom)
\begin{equation}
E = \pm e/R^2.
\end{equation}
The expressions (11) and (12) are used in the volume integral of the
equation (6). The integration will be carried out on the effective
volume of Salam's barrier which is defined by the effective radius
$r_{eff}$. The account of the independence of the coordinates $R$
and $r_{eff}$ results in
\begin{equation}
(m - m_{cc})c^2 = (1/6)(r_{eff}/R)^3e^2/R.
\end{equation}
The right hand side of the last equation grows up due to account of
the central interaction. The term $m_{cc}$ in the left hand side of
(13) is the varying quantity and implicitly characterizes the rest
mass which has been left at the particle after the compensation of
the central and non-central interactions.
\par The expression (13) shows that the location of a state with respect
to the curve $e^2/R$ (in the considered system it represents the curve
-$E_p$) is defined by the ratio $r_{eff}/R$. The double physical meaning
is attributed to this value in the model. On the one hand, it is connected
with the curvature of the spatial representation of Salam's barrier.
On the other hand, it characterizes the distribution of the barrier
along the generalized coordinate $R$ (interaction radius) and consequently
can be connected with the wave function in its probability interpretation.
The gauge transformations arise in this aspect as a result of the
periodic reccurence of the states with the same value of the ratio
$r_{eff}/R$. The characteristic properties of the variations of this
quantity along the gauge cycles are considered below for the confirmation
of this statement.
\par It was mentioned in the section 6 that one cycle begins and ends
by the states which are located on the curve $-E_p$. It is
\begin{equation}
-E_p = e^2/R.
\end{equation}
in the considered system.
It follows from the expressions (13) and (14) that the beginning and end
of the cycle correspond to the same value
\begin{equation}
r_{eff}/R = 6^{1/3}
\end{equation}
\par The numerical values of the ratio $r_{eff}/R$ in other states
can be determined with the account of the properties 1-3 of the section
6. The states which are located bellow $e^2/R$ are marked by the subscript
$e$ $(r_e/R)$, and the states higher - by the subscript $p$ $(r_p/R)$.
The system from two equations arises in this case
\begin{equation}
\left \{
\begin{array}{ l l }
(m - m_{cc})c^2 = (1/6)(r_e/R)^3e^2/R \ ; \ \ \ & \nonumber \\
(M - M_{cc})c^2 = (1/6)(r_p/R)^3e^2/R \ . \ \ \ &  \\
\end{array}
\right.
\end{equation}
\par The division of the second equation on the first one results
in the expression
\begin{equation}
\frac{M - M_{cc}}{m - m_{cc}} = \frac {(r_p/R)^3}{(r_e/R)^3},
\end{equation}
which connects the effective values of the rest masses to the appropriate
values of the ratios $r_e/R$ and $r_p/R$. The expression (17) defines
the ratio of the rest masses $M/m$ in the limiting borders of Salam's
range.
\par Two cycles $aR_o1d'$ and $d'd''b''f$ represent the first and second
cycles of the gauge transformations in the figure 1.
\par The state $R_o$ in the first cycle is located on the longitudinal
axis that is zero value of barrier volume should be attributed to
it. However, it is necessary to note that it is connected with the
conventional location of the quantity $e^2/R$ on infinity. Actually
the  nonvanishing value of this quantity should be attributed to this
state.
\par The values $r_e/R$ and $r_p/R$ are not arbitrary. The connection
between them is defined by the condition of the location of the point
$a'$ on the curve $e^2/R$. This point corresponds to the termination
of the term $mv^2/2$ in the series (3) and the begining of the term
$(3/8)mv^4/c^2$ in the series (4). The point $a'$ is the common point
of these two segments in this aspect.
\par It follows from the first and second equation (16) that the
common point should satisfy to the condition
\begin{equation}
r_e/R = r_p/R = 6^{1/3}.
\end{equation}
\par By multiplying of the first equation (16) into the coordinate
$r_p$, one can get equation connecting the ratios $r_e/R$ and $r_p/R$
for any state
\begin{equation}
(r_e/R)^3r_p/R = A,
\end{equation}
where
\begin{equation}
A \equiv \frac {6(m - m_{cc})c^2r_p}{e^2}.
\end{equation}
One can see from the equation (19) that the condition of the equality
$$
r_e/R = r_p/R
\nonumber
$$
is satisfied for different values of $A$ but the condition (18) is
satisfied only in one case when the common point belong also to the
curve $e^2/R$. In other cases the common point lays above or below this
curve. It follows from here that the considered values are connected
among themselves by the relation
\begin{equation}
(r_e/R)^3(r_p/R) = 6^{4/3}.
\end{equation}
\par One has to note that any state has two points of the crossing with
the curve $e^2/R$ - in the longitudinal and lateral directions. The
parity (21) connects events both in one cycle (lateral transitions) and
in consequtive cycles (longitudinal transitions) due to this fact.
Analogous relation for the longitudinal transitions is
\begin{equation}
(r/R_e)^3(r/R_p) = 6^{4/3},
\end{equation}
where $R_e$ refer to area I and $R_p$ - to area II in the figure 1.
\par It follows from (21) and (22) that
\begin{equation}
\left \{
\begin{array}{ l l }
r_e/R = r/R_e \ ; \ \ \ & \nonumber \\
r_p/R = r/R_p \  \ \ \ &  \\
\end{array}
\right.
\end{equation}
for all gauge cycles. Two layers are formed in Salam's range side
by side with the layer $-E_p$ due to the relation (23). One of them
defines the movement of the particle $p$ in the range and the other
- of the particle $e$. These layers are distinguished by the various
values of $r/R$.
\par One can define the values $r_e/R$ and $r_p/R$ at the unlocal
representation of the barrier of the particle which is placed in the
layer $r_p/R$. Such representation is based on the distinction of
the physical properties of three layers forming the gauge cycle. The
distinction arises at the point particle approximation.
\par The location of the point particle in the layer $-E_p$ is connected
with the geometrization principle. Since the terms quantity of the
series (4) defines the variation of the nonrelativistic kinetic energy
$E_k$ relative to $R_{\infty}$, the beginning and end of any transition
have to belong to the curve $-E_p$. In the considered system $r_e/R
+ r_p/R$ both particles are not located on $-E_p$. Therefore, the
geometrization principle can not be satisfied at the transitions of
the point particle between layers $r_e$ and $r_p$. At the same time,
the location in this layers has the property which is important too.
This property can be defined at the unlocal consideration which is
connected with the model role of the electrical charge.
\par The role of the charge is connected with the transformation of the
lateral segments into the longitudinal ones. The scale of this transformation
is defined by the quantity $-E_p$ depending on the value of the charge.
Therefore, the charge defines the compression of the field energy on
the longitudinal direction on the contrary to the rest mass which
defines the compression of the energy on the lateral one.
\par The definition of the charge role allows to define such transitions
at which both the geometrization principle and property of the layers
$r_e/R$ and $r_p/R$ are valid. The satisfaction of the both properties
can be completed in this case when the particle has the part of the
whole charge $+e$ in the state $r_p/R$. Then the state $r_e/R + r_p/R$
can be considered as the interaction of the particle which has the charge
$-e$ with another particle having fractional charge. If the value of the
fractional charge is agreed with the structure of the gauge cycle in the
longitudinal direction, the whole charge $+e$ corresponds to one gauge
cycle. Therefore, the interaction of two point particles with charges
$\pm e$ can be considered as the interaction of the particle which
localized at the beginning of the cycle with another particle which
is located at the end of one. Such unlocal representation arises due
to the space structure of the Salam's barrier. This structure has
being agreed with gauge cycle one. Therefore, the summary structure
of the transition between adjacent states is the superposition of the
gauge cycle and the structure corresponding to the transitions between
the layers $r_e/R$ and $r_p/R$. The quantity of such transitions in
one gauge cycle has to correspond to the whole charge $+e$. Then the
fractional charge $q$ can be represented as
\begin{equation}
q = + e/l
\end{equation}
due to the functional role of the charge ($l$ is the transitions quantity
in one cycle). The last relation represents average charge connecting
to quantity of the transitions between the layers $r_e/R$ and $r_p/R$
in one cycle.
\par The transitions quantity $l$ takes the integer values. The case of
three transitions is of special interest. It is connected with quantity
of the transitions which is equal to three according to the results
of the previous section. If the quantity of the transitions between
the layers $r_e/R$ and $r_p/R$ is also equal to three, then the sequence
of the stationary states has the harmonic character connecting with
the periodical reiteration of the all values of $r/R$ in one cycle.
\par Two states $R_o1$ and $d''b''$ represent two adjacent states
of the system $r_e/R + r_p/R$ in the figure 1. The state $R_o$ corresponds
to $r_e/R$ and the state $1$ - to $r_p/R$ in the first gauge cycle.
Analogously, the state $d''$ corresponds to $r_e/R$ and the state $b''$
- to $r_p/R$ in the second one. One can see from the figure 1 that
the quantity of the gauge transitions (connected with the crossing
of the curve $-E_p$) between these states is equal to two ($R_o1 \rightarrow
1d''\rightarrow d''b''$). According to the results of the previous
section, such quantity is equivalent to the rotation of the system
on the angle $2\pi$. The double transition can not lead to the harmonic
sequence. It has to be equal to three ($l = 3$) in order to satisfy
to this condition. However, it is impossible to pass from the space
with index $e$ (or $p$) to the space with the same index with help
of three transitions between layers $r_e/R$ and $r_p/R$. This property
is connected with the general property of the spinor objects - the
rotation of the system on an angle $2\pi$ does not lead to the initial
state [8]. The additional rotation on the same angle $2\pi$ is necessary
due to this reason. The additional rotation on the same angle means
the transition to the position of the end of the second gauge cycle.
Finally, one can see that the system $r_e/R + r_p/R$ comes into the
initial state with help of two gauge cycles. Therefore, the value
of $l$ in the relation (24) is equal to six and the mean charge
\begin{equation}
q = + e/6.
\end{equation}
\par Thus, the system $r_e/R + r_p/R$ can be treated as the interaction
of the particle $-e$ with the particle $+e/6$ in the aspect of the
gauge cycles. Then, the state of the first particle has to be decreased
into six time in comparison with the state predicted by the curve $-E_p$.
Then, it follows from the first equation of the system (16) that
\begin{equation}
r_e/R = 1.
\end{equation}
\par By using of the last expression in the equation (21), one can
obtain
\begin{equation}
r_p/R = 6^{4/3}.
\end{equation}
\par It is necessary to pay attention to the physical meaning of the
relation (26). The interaction area is a sphere with the radius $R$
at the central interaction. Then, the equality of the radii $R$ and
$r$ means that the influence of the external field is compensated
by Salam's barrier in the interaction area. Therefore, the surface
integral in the equation (1) is equal to zero and the stationary equation
(6) is satisfied. If the quantity $l$ is not equal to six, the value of
$r_e/R$ is changed. Then the influence of the external field is not
compensated by Salam's barrier and the equation (6) is not satisfied.
Thus, the harmonic sequence of the stationary states, which is connected
with the periodical reiteration of the all values of $r/R$ in one
gauge cycle, can be executed at one value $l = 6$.
\par The relation (27) has the important meaning together with (26).
The mutual action of these relations allows to understand the role
of the unlocal interaction of two particles. It is connected with
two conditions of the gauge cycles - the geometrization principle
and the necessity of the compensation of the external fied influence
in the interaction area. The first one is executed at location of 
the point particle on the curve $-E_p$. The second one is satisfied
in the case of the location of such particle in the system $r_e/R
+ r_p/R$. The mutual execution of these conditions is possible only
if Salam's barrier of the particle $p$ is unlocal. At such representation
it possesses the both necessary conditions. One can see that this
fact is connected with dualistic representation "wave-corpuscle" of
the elementary particle.
\par The unlocal structure of the barrier defines the internal area
of the coordinate $R$ in any stationary state. The interaction of
the fractional electrical charge with charge $-e$ is essential in
this area. Therefore, it corresponds to the strong interaction. Thus,
the whole area of the coordinate $R$ variations is divided into three
parts in two-particles interaction. The external part corresponds
to the variation of the nonrelativistic kinetic energy $E_k$ (relative
to $R_\infty$) in the right hand side from the fixed value $R_i$.
The spacing of the double gauge cycles is important in this part.
The second part is connected with the quantity (relative to $R_{\infty}$)
of the terms of the series (4) at $R_i$. It defines the energy stock
which will be transformed into $E_k$ when the coordinate $R$ decreases
relative to $R_i$ in the one-particle approximation. The space of
the transformation defines the third part of the area. The last one
is the area of the strong interaction in the two-particles approximation.
\par The quantity of the terms of the series (4) decreases as $R$
decreases. The limit is the case when the quantity of the terms
is equal to zero and the whole stock of the potential energy of the
series (4) is exhausted. This limit is represented by the state $R_c$
in the figure 1. The presence of this limit is important property
of the model. Its calculated value is about 5 Fm which is in agreement
with the radius of nucleus. Since the point particle $p$ is located
at the end of the gauge cycle and the particle $e$ - at its beginning,
zero value of the quantity of the terms for the particle $p$ corresponds
to unity value for the particle $e$. This property leads to unity
value of the main quantum number for the particle $e$ in the end.
\par In fact the limiting state corresponds to the core of nucleus.
Since the particle $p$ is located at the end of any gauge cycle, the
core is presented in any one. The value of the coordinate $R$ (corresponding
to the core) decreases as $R$ decreases. Thus any state defines two
volumes of Salam's barrier. The first one, corresponding to the location 
of the particle $e$, defines the maximal volume. The second one, corresponding 
to the localization of the particle $p$, defines the minimal volume. 
Therefore, the process of the coordinate $R$ decrease can be considered 
as consecutive process of the transitions between maximal and minimal 
values of Salam's barrier.
\par The unlocal structure of the Salam's barrier leads to variation
of the mass in the lateral direction analogously to the longitudinal
variation of the charge. Such variation is connected with the correction
factor in the relation (17). The value of this factor is connected
with the variation of the angular momentum.
\par It follows from the (16)
\begin{equation}
\left \{
\begin{array}{ l l }
(m - m_{cc})cR = (1/6)(r_e/R)^3e^2/c; &  \nonumber \\
(M - M_{cc})cR = (1/6)(r_p/R)^3e^2/c,  &   \\
\end{array}
\right.
\end{equation}
which shows that the angular moments $(m - m_{cc})cR$ and $(M - M_{cc})cR$
do not depend on the value $R$ and are constant in the all range of
changes of $R$. The variation of these moments $\Delta P$ only occurs
with transition between layers $r_e/R$ and $r_p/R$. It is necessary
to distinguish the transitions between full gauge cycles (which correspond
to two gauge cycles and full charge $+e$) and the transitions in the inner
area of the gauge cycles (which correspond to fractional charge $q$).
\par The interior transitions are connected with the variation of the
charge in the longitudinal direction and with the quantity of the ratio
$(M - M_{cc}/(m - m_{cc})$ in the lateral one. Since the values of $r_e/R$
and $r_p/R$ do not change in whole Salam's range, the variation $\Delta P$
at one transition between layers is constant and the total variation
$\Delta P_l$ at $l$ transitions is
\begin{equation}
\Delta P_l = l\Delta P.
\end{equation}
\par The total quantity of the number interior transitions $l$ correspond
to net charge $+e$. Then the relation (29) shows that $\Delta P$
is compared to the some value of the elementary fractional charge
which is repeated in all interior transitions at the longitudinal
direction. The some elementary value of the ratio $(M - M_{cc})_e/(m - m_{cc})$
(which is repeated at the lateral direction) is present analogously
to the fractional charge. Then the summary variation $(M - M_{cc})_l/(m - m_{cc})$
is
\begin{equation}
(M - M_{cc})_l/(m - m_{cc}) = l(M - M_{cc})_e/(m - m_{cc}).
\end{equation}
\par If one does not take into account the unlocal structure of Salam's
barrier, the quantity of the transitions between states $r_e/R + r_p$
in full gauge cycle ( two gauge cycles) is equal to four. In this
case
\begin{equation}
(M - M_{cc})_l/(m - m_{cc}) = 4(M - M_{cc})/(m - m_{cc}),
\end{equation}
where $(M - M_{cc})/(m - m_{cc})$ corresponds to the value which is
predicted by the relation (17).
\par It follows from (30) and (31)
\begin{equation}
(M - M_{cc})/(m - m_{cc}) = (l/4)(M - M_{cc})_e/(m - m_{cc}).
\end{equation}
\par The relation (32) shows that the correction factor in the ratio (17)
is equal to $l/4$ and this one is
\begin{equation}
(M - M_{cc})/(m - m_{cc}) = (l/4)(r_p/R)^3/(r_e/R)^3.
\end{equation}
\par The substitution of $l = 6$, the values (26) and (27) into the
last relation leads to
\begin{equation}
(M - M_{cc})/(m - m_{cc}) = 1944
\end{equation}
which will be satisfactory coordinated with the ratio of proton and
electron rest masses. The agreement can be even better if one takes
into account that the difference $\sim$50 MeV can be treated as total
mass defect which can be spent on connection energy between nucleons.
\par Since the quantity (34) corresponds to the only value of $l$
satisfying to (26) and (27), the ratio of masses (34) defines the
only particle $p$ which can form the stationary pair with the particle
$e$. If the ratio is different from (34), the geometrization principle
is satisfied but the condition (26) is violated. Therefore, such pairs
can only form the quasi-stationary states which have not the harmonic
reiteration (for example, the systems $ {\mu}^+ + e^-$ and $e^+ + e^-$).
\par The important confirmation of the model correctness gives numerical
definition of the variation of the angular momentum $\Delta P$ at one
transition between layers $r_e/R$ and $r_p/R$. It is equal to
\begin{equation}
\Delta P = (1/6)[(r_p/R)^3 - (r_e/R)^3]e^2/c
\end{equation}
according to relation (28). Inserting of the numbers of fundamental
constants $e$ = 4.803 $\cdot10^{-10}CGSE$, $c$ = 2.998 $\cdot10^{10}cm/sec$,
the relations (26) and (27) into the last expression gives
\begin{equation}
\Delta P = 1.661\cdot10^{-27}erg/sec.
\end{equation}
The last value coincides accurate up to second sign after point with
\begin{equation}
\Delta P = (\pi /2)\hbar,
\end{equation}
where $\hbar$ is Plank's constant ($\hbar$ = 1.054$\cdot10^{-27} erg/sec$).
If one takes into account that two gauge cycles correspond to four
transitions between the layers (26) and (27) in the external area of $R$
variations, then the summary variation $\Delta P_{gc}$ is
\begin{equation}
\Delta P_{gc} = 2\pi\hbar
\end{equation}
in this area. The last expression shows good agreement not only in
value of $\hbar$, but also in angular correlations.
\par It is necessary to underline that the equality (38) takes place
for the stationary pair of the particles with the ratio of the rest masses
being in agreement with the relation (34).
\par The important confirmation of model correctness gives also numerical
definition of the rest mass variation at transition between the adjacent
pairs of the gauge cycles. It is necessary to take into account the
correction factor $l/4$ in the expression (33) in such procedure.
The constant of the gravitational interaction arises at the definition.
\par One can obtain with help of the expressions (21) and (22) the variation
of the coordinate $R$ corresponding to two gauge cycles
\begin{equation}
R_k = \frac{(r_p/R)^6}{(r_e/R)^6}R_{k+2},
\end{equation}
where
\begin{equation}
R_{k+2} < R_k < R_{\infty}
\end{equation}
and $R_k \rightarrow R_{k+2}$ means the transition at two gauge cycles.
\par The first equation (16) leads to the expression
\begin{equation}
(m - m_{cc})_kc^2 = (1/6)(4/l)(r_e/R)^3e^2/R_k.
\end{equation}
for the state $k$ (with account of the correction factor $l/4$ in the
expression (33)) relative to the state $k + 2$.
The substitution of the relation (39) into the expression (41) with account of
the equality (26) gives
\begin{equation}
(m - m_{cc})_kc^2 = \frac {\gamma e^2}{R_{k+2}},
\end{equation}
where
\begin{equation}
\gamma = \frac {1}{6\cdot1.5\cdot6^8} = 6.61\cdot10 ^{-8}.
\end{equation}
\par The value (43) coincides accurate up to second sign after point
with the constant of the gravitational interaction. It defines value
$(m - m_{cc})_kc^2$ with decreasing $R$ on value corresponding to
two gauge cycles. The coincidence shows that the potential stock of
the energy $E_{pl}$ at any space point can be considered as the stock
of the energy of the gravitational interaction. According to relations
(4) and (10), it is defined by
\begin{equation}
E_{pl} = - (3/8) mv^4/c^2 - (5/16) mv^6/c4 - ...
\end{equation}
\par Such definition means that the relation (9) corresponds to
\begin{equation}
mv^2 = mc^2 + E_{pl},
\end{equation}
which shows that the growth of the double quantity $E_k$ occurs due
to the decrease of the potential energy $E_{pl}$ of the gravitational
interaction.
\par Thus, the interaction with the potential energy $E_{pl}$ defines
three fundamental interactions - gravitational, electromagnetic and
strong ones. It divides the whole area of the variations of the coordinate
$R$ changes into three parts. The external part (in the right hand
side from the considered value $R_i$) corresponds to the electromagnetic
interaction. It is connected with the variation of the nonrelativistic
kinetic energy of the particle $e$ in the longitudinal direction.
The second part corresponds to the stock of the gravitational energy
in the lateral direction. The third part corresponds to the internal
area of Salam's barrier of the particle $p$. The interaction of the
fractional electrical charge with charge $-e$ is significant in this
area. Therefore, it corresponds to the strong interaction.
\par It is necessary to underline, that the value (43) is obtained
by use of the value of $r_e/R = 1$ in the relation (41). The last
one, only, corresponds to the interaction  of the particle $e$ with
the particle $p$, rest mass of which is determined by the ratio (34).
If the rest mass of the one is different (the quasi-stationary pairs),
the value of $\gamma$ differs from the experimental value of the constant
of the gravitatational interaction. It means, that the experimental
value of the gravitational constant corresponds to the gravitational
field which is formed in the interaction of the electron with the
proton. This conclusion can be of large interest. When quasi-stationary
particles (which can form the quasi-stationary pairs with the electrons)
hit earthly atmosphere (for example, the positrons at the higher sun
activity) the areas of local infringements of the gravitational field
with accompanying physical effects can arise. This areas can serve
as the sources of the gravitational waves.
\par It is also necessary to underline another role of the particle
$e$. The interaction with it defines the stationary and quasi-stationary
pairs of the particles for Salam's range which is defined by the rest
mass of the particle $e$. These pairs differ by values of the quantum
number $l$ which is defined with the quantity of the transitions
between layers $r_e/R$ and $r_p/R$. The possibility of the classification
of the elementary particles, belonging to such range, arises there.

\section{CONCLUSION}

\par 1. The gauge model of the fundamental interactions unification
is offered. The model takes into account the influence of the effect
of the rest mass compensation (Salam's effect) in the field equation
of the energy conservation. The stationary solutions of this equation
are considered. The condition of the stationarity is ensured by the
Salam's barrier which is formed due to the transformation of the rest
mass of the particle into it's energy. The barrier compensates
the influence of the external field in the considered volume. The
formal condition of the compensation is the equality of the radii
$R$ and $r$ in the spherical approximation. The radius $R$ defines
the area of the interaction. This area is a sphere at the central
interaction. The radius $r$ defines the effective volume of Salam's
barrier which compensates the influence of the external field. The
condition of the equality of the radii is the first from two considered
conditions.
\par 2. The second condition is connected with the expansion of the
relativistic kinetic energy into McLoren's series. The nonrelativistic
kinetic energy $E_k$ is the second term of this expansion. The geometrization
principle (variations of the field potential energy should be equal
to variations of the $E_k$) is used for the separation of the influence
of $E_k$ from the influence of the sum of the terms of $E_{r3}$ with
higher degrees. The execution of the geometrization principle is the
second condition of the model. It is shown, that the action of Salam's
effect results in the separation of the influence of the coordinates
$R$ and $r$. The influence of $R$ is connected with the longitudinal
direction and the influence of $r$ - with the lateral one. The sum
$E_{r3}$ defines, at the lateral one, the stock of the field
energy which is transformated into $E_k$ at the decrease of $R$ at
the longitudinal one. The area of the variations of $R$ is divided
into three parts due to this property. These parts are connected with
the arrangement of the considered value of $R_i$ relative to $R_{\infty}$.
The first part is connected with the value of $E_k$ relative to $R_{\infty}$.
It is formed due to the decrease of the terms quantity (relative to
$R_{\infty}$) in $E_{r3}$. The second part corresponds to the quantity
of this terms remaining in $E_{r3}$ (relative to $R_{\infty}$) in
the lateral direction. The third part corresponds to the transformation
of the remaining quantity of the terms into $E_k$ in the process of
the decrease $R < R_i$.
\par 3. Three parts of the variation of $R$ lead to the gauge cycle
consisting of three gauge transitions. Two of them correspond to
the longitudinal segments defining the transformation of $E_{r3}$
to $E_k$ in the right and left hand sides from $R_i$. The third transition
corresponds to the quantity of the transformated parts of $E_{r3}$
in the lateral direction. The quantities of the gauge transitions
are connected with three functional dependences from the generalized
coordinate $R$. The first one represents the dependence of the field
energy $-E_p$ which is equal to the potential energy on the absolute
quantity, but has different sign in the connected systems. According
to the geometrization principle, it defines the beginning and end
of the gauge cycle. The second and third dependences are similar
to the curve $-E_p$ and differ from it by the similarity coefficient.
They define beginning and end of the lateral segment corresponding
to the transformated parts of $E_{r3}$.
\par 4. Three mentioned above curves differ from each other by the
similarity coefficient. It is shown, that this one is defined by the
ratio $r/R$ at Coulomb interaction of two particles with electrical
charge $\pm e$. The double physical meaning is attributed to this
ratio in the model. On the one hand, it is connected with the curvature
of the spatial representation of Salam's barrier. On the other hand,
it characterizes the distribution of the barrier along the generalized
coordinate $R$ (interaction radius) and consequently can be connected
with the wave function in its probability interpretation. The gauge
transformations arise in this aspect as a result of the periodic reccurence
of the states with the same values of the ratio $r/R$. Three dependences
which were discussed above correspond to three layers with different
values of $r/R$.
\par 5. The character of the reiteration of all values of $r/R$ is
important at the transitions between adjacent gauge cycles. The beginning
and end of the gauge cycle always correspond to the same value of
$r/R = 6^{1/3}$. This value corresponds to the curve $-E_p$ and is
connected with the second condition of the model (execution of the
geometrization principle). The quantities $r_e/R$ and $r_p/R$, corresponding
to two other functional dependences of the gauge cycle, are connected
with the first condition of the model (the necessity of the compensation
of the external field influence by the Salam's barrier in the area of
the interaction). These quantities are connected with the internal
structure of the gauge cycle. The beginning and end of the cycle are
connected with each other by three gauge transitions. At the same
time, the quantity of such transitions is equal to two at transitions
between adjacent states $r_e/R + r_p/R$. Therefore, such transition
can not be harmonic. The quantity of the gauge transitions have also
to be equal to three at the transitions between adjacent states $r_e/R
+ r_p/R$. Then, the sequence of the stationary states have the harmonic
character. The variation of the quantity of the gauge transitions
in the interior of the gauge cycle divides the area of the coordinate
$R$ variation into the internal and external parts. The variations of the angular
momentum $[m - m_{cc}]cR$ and ($[M - M_{cc}cR])$ correspond to whole
charge $+e$ at the interaction $\pm e$ in the external area. The variations
of these momenta correspond to the interaction of the charge $-e$
with fractional charge $+e/l$ in the internal area ($l$ - the quantity
of the gauge transitions between adjacent states $r_e/R + r_p/R$).
The variations of the charge occur on the longitudinal direction.
The variations of the ratio $(M - M_{cc})/(m - m_{cc})$ occur on
the lateral one. The variations of this ratio and of the charge
correspond to the complex structure of Salam's barrier around the
particle which is become localized in the layer $p$.
\par It is shown that the harmonic reiteration of all values $r/R$ (in
one gauge cycle) corresponds to values of $r_e/R = 1$ and $r_p/R =
6^{4/3}$. The first relation shows that Salam's barrier compensates
the influence of the external field in the area of the interaction
(the sphere with the radius $R$) of the particle $e$ (which become
localized in the layer $r_e/R$) with the particle $p$ (which become
localized in the layer $r_p/R$). Therefore, the first condition of
the model is only executed in case of the harmonic reiteration of
all values of $r/R$. If the reiteration at transitions between adjacent
states $r_e/R + r_p/R$ is not harmonic, this condition is broken.
\par The whole charge $+e$ of the particle $p$ corresponds to execution
of both conditions in case of the harmonic reiteration. The states
of the particle $-e$ are stationary at the interaction with the whole
charge $+e$ (external area of the gauge cycle) due to this property.
The external area corresponds to the electromagnetic interaction.
As the internal area of the gauge cycle corresponds to the interaction
of the charge $-e$ with fractional charge $+e/l$, this one corresponds
to the strong interaction.
\par The agreement between calculated values of the variations of the
angular momentum and the ratio of the rest masses of the proton and
electron with their experimental values confirms the important role
of two gauge conditions in the unification of the fundamental interactions.
\par 6. The potential energy $E_{pl}$ defined by the relations (44)
plays the special role in the unification mechanism. The agreement
of the numerical definition of the variation of the rest mass at the
transition between adjacent stationary states with the constant of
the graviational interaction shows that this energy has the meaning
of the potential energy of the gravitational interaction.
\par The energy $E_{pl}$ divides the whole space of the variations
of the coordinate $R$ to external and internal parts at any value
of $R_i$. The electromagnetic interaction is predominant in the external
part. The internal part is the area of the strong interaction. The
gravitational interaction raises the electromagnetic and strong interactions
in this aspect.
\par The energy $E_{pl}$ has the important property in the model.
Since the quantity of the terms in the series (4) decrease at the
consecutive decrease of the coordinate $R$, the lower limit (of
the stationary interaction of the particle $-e$ with the particle
$+e$) exists. This formal limit corresponds to one term (relative
to $R_{\infty}$) in the series (4).
\par The preliminary results of this work are submitted in [9-11].

\section{REFERENCES}

1. A. Salam, Proc. XVIII Int. Conf. on Nucl. Phys., Tbilisy (1976),
   D1-2-10400, Dubna, v.2, (1977), 91 \\
2. L.D. Landau, E.M. Livshitz, Field Theory, Moscow, (1962), 97 \\
3. H. Weyl, I.Z.Phys., v.56, (1929), 330-352 \\
4. N. Rozen, J.Phys.Rev.,v.57,(1940), 147-153 \\
5. A. Papapetrou, Proc.Roy.Irish Acad.,v.A52, (1948), 11-23 \\
6. S. Gupta, Proc.Phys.Soc.,v.A65, (1952), 608-619 \\
7. W. Thirring, Ann.of Phys.,v.16, (1961), 96-117 \\
8. R. Penrose, W. Rindler, Spinors and space-time, Moscow,"World", v. 1,(1987), 62 \\
9. R. Ragachurin, "Non-linear relativistic model charged elementary
                  particle", Preprints INP (Tashkent), P7-208, v.1,
                  (1986), 3-14; P7-243, v.2, (1986), 3-24; P7-293, v.3, (1987), 3-38;
                  P7-352, v.4, (1988), 3-46 \\
10. R. Ragachurin, FB16 Book of Abstract, ed. by Chi-Yee Cheung and
                  Shin-Nan Yang, Taipei (2000), 241-242 \\
11. R.G. Ragachurin, The Book of Abstracts of LI Meeting on Nuclear
                     Spectroskopy and Nuclear Structure, Sarov, (2001), 
65-66

\clearpage
\thispagestyle{empty}

\begin{figure}
\rotate{\epsfbox{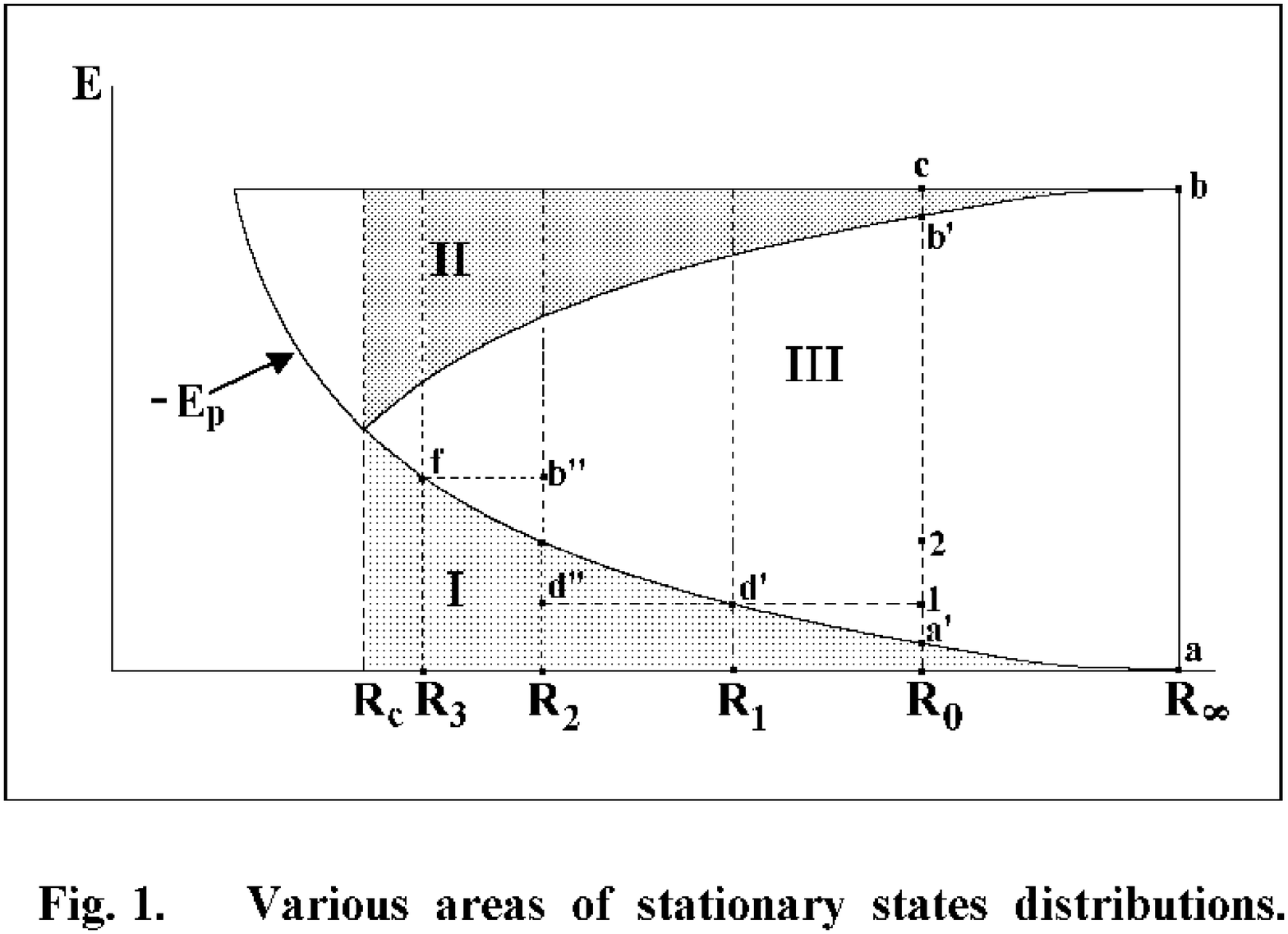}}
\end{figure}

\end{document}